\documentclass[prl,aps,twocolumn,groupedaddress,floats,showpacs,final]{revtex4-1}
\usepackage{graphicx}
\usepackage{dcolumn}
\usepackage{bm}
\usepackage{color}
\definecolor{blue}{rgb}{0.3,0.3,0.9}

\begin{document}

\title{The Higgs mode in a two-dimensional superfluid}

\author {L. Pollet$^{1}$ and N. Prokof'ev$^{1,2,3}$ }

\affiliation {$^1$Department of Physics and Arnold Sommerfeld Center for Theoretical Physics, Ludwig-Maximilians-Universit{\"a}t M{\"u}nchen, D-80333 M{\"u}nchen, Germany}
\affiliation {$^2$Department of Physics, University of Massachusetts, Amherst, Massahusetts 01003, USA}
\affiliation{$^3$Russian Research Center ``Kurchatov Institute", 123182 Moscow, Russia
}
\date{\today}

\begin{abstract}
We present solid evidence for the existence of a well-defined Higgs amplitude mode
in two-dimensional relativistic field theories based on analytically continued results from quantum Monte Carlo
simulations of the Bose-Hubbard model in the vicinity of the superfluid-Mott insulator
quantum critical point, featuring emergent particle-hole symmetry and Lorentz-invariance.
The Higgs boson, seen as a well-defined low-frequency resonance in the spectral density,
is quickly pushed to high energies in the superfluid phase and disappears by merging with the
broad secondary peak at the characteristic interaction scale.
Simulations of a trapped system of ultra-cold $^{87}$Rb atoms
demonstrate that the low-frequency resonance 
is lost for typical experimental parameters, while the characteristic frequency for the onset of strong response is preserved.
\end{abstract}


\maketitle
The emergence of low-energy excitations in systems with spontaneously broken symmetry is one of the most
fascinating and fundamental subjects in physics relevant for understanding such diverse systems
as solids, magnets, ultra-cold atoms, and relativistic fields.
The generation of mass by the Anderson-Higgs mechanism \cite{anderson,higgs} is particularly important
for the Standard model \cite{standard}, where detection of the Higgs boson is still the missing link in revealing this mechanism,
as well as for numerous superfluid/superconducting condensed-matter 
systems.
In realistic materials the amplitude mode is often masked by other low-energy excitations.
These complications are avoided by considering atomic bosonic superfluids, described by a
complex order parameter field, which constitute the cleanest experimental realization.

Generic superfluids do not feature a well-defined Higgs boson
(by 'well-defined' we understand a mode seen as a sharp resonance).
Weakly-interacting gases do not have it because 
at and around the critical temperature $T_c$ for the superfluid (SF) to normal fluid phase transition
all long-wave elementary excitations are overdamped,  while at $T \to 0$ the low-energy spectrum
is exhausted by the Bogoliubov quasiparticle excitations where phase and density 
are canonical variables.
Strong interactions do not necessarily change this picture.
As long as the critical temperature remains large, as in $^4$He,
long-wave excitations are overdamped at $|T-T_c| \ll  T_c$.
At low temperature,  only the Nambu-Goldstone
phase modes remain at low frequencies whereas the amplitude mode is pushed to the incoherent continuum at
the (large) characteristic interaction energy scale. Suppressing $T_c$ by increasing interactions
may trigger a first order transition to the solid phase and not work either.
It is thus crucial to consider an experimental system with a second-order quantum critical
point (QCP) where $T_c$ for superfluidity can be tuned to near zero.

The Bose-Hubbard model 
\begin{equation}\label{BH}
H = - J \sum_{<ij>}  b_i^\dag b_j^{\,}
+ \frac{U}{2} \sum_i n_i(n_i-1) - \sum_i (\mu -v_i)n_i  \, ,
\end{equation}
with experimentally adjustable ratios between the hopping amplitude $J$, on-site
interaction $U$, chemical potential $\mu$, and trapping potential $v_i$,
provides an accurate description of ultra-cold bosonic atoms in optical lattices. 
At integer filling factor, $\nu = \langle n_i \rangle$, and zero temperature
it undergoes a second-order quantum phase transition from SF to the Mott insulator
(MI) phase as the interaction strength is increased \cite{fisher89}. The critical field
theory behind this transition is Lorentz-invariant and particle-hole symmetric (while the SF-MI transition 
for generic values of $\nu$ belongs to the universality class of the ideal Bose
gas at vanishing density and is excluded)~\cite{fisher89,book}. Despite the decay into two
phase modes the existence of a sharp Higgs boson is guaranteed in
two special limits: (i) in a three-dimensional (3D) system where the corresponding 4D quantum field
theory is at the upper critical dimension with asymptotically exact mean-field behavior
and vanishing decay rates (see Ref.~\cite{ruegg} for an experimental observation
in a quantum antiferromagnet);
(ii) at large momentum when the relativistic time dilation effect leads to an increased 
quasiparticle decay time. 
%
\begin{figure}[htbp]
\vspace*{-0.4cm}
\hspace*{-1.5cm} \includegraphics[angle=0,width=1.\columnwidth]{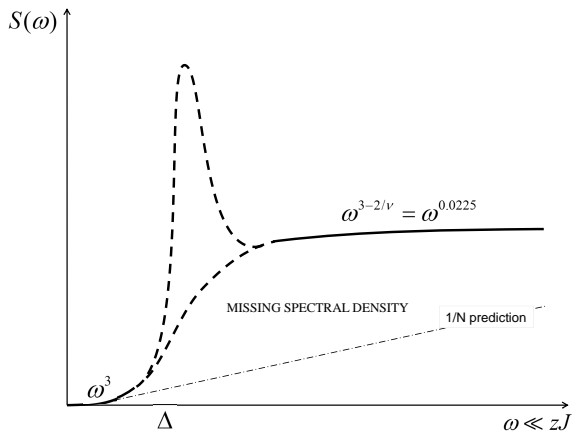}
\vspace*{-0.5cm}
\caption{\label{fig:expectations} Universal scaling predictions for the scalar susceptibility
(solid lines). The dashed-dotted line depicts prediction of Ref.~\cite{pod11} and misses most of the spectral density at the relevant energy scale $\Delta \propto (1-U/U_c)^{\nu}$.
The two alternatives for connecting universal power laws are shown by dashed lines
(one may also imagine multiple peaks in the crossover region).
 }
\end{figure}

The most intriguing question is whether the low-frequency Higgs boson can be seen as a
well-defined excitation at zero momentum at the density-driven QCP
of the 2D Bose-Hubbard model and how it disappears with detuning to the SF phase.
Equally important are finite temperature effects and the
role of the trapping potential in experiments.
A theoretical treatment of the Higgs amplitude mode is notoriously difficult and controversial.
In Refs.~\cite{chubukov,sachdev,book,zwerger} exact scaling laws in the low-frequency limit
were established, as well as arguments given that the mode is at the edge of the two-phonon continuum,
rendering the mode overdamped.
Huber {\it et al.} used a variational Ansatz which, however, predicted a spurious first order SF-MI transition, and thus
was limited to the parameter regime away from quantum criticality~\cite{huber, huber2}.
Podolsky {\it et al.} generalized the field theoretical results of Ref.~\cite{chubukov} to high frequencies
and discussed in detail the response function for the order-parameter density (scalar response) 
within a $1/N$ and a weak coupling expansion schemes~\cite{pod11}.
They revealed a broad peak whose maximum saturates at finite value at the QCP and concluded that
close enough to the transition, it becomes impossible to identify the Higgs energy with the peak position. 
Their findings are in quantitative and qualitative disagreement with those reported here, as is detailed in the supplementary material~\cite{epaps}.
The major problem with the results of Ref.~\cite{pod11} is the strong violation of the universal low-frequency
scaling law for the scalar response function \cite{book}, $S(\omega )\propto \Delta^{3-2/\nu} F(\omega /\Delta )$, where
$\Delta \propto (1-U/U_c)^{\nu}$ is the characteristic energy scale in proximity of the quantum critical
point, and $\nu = 0.6717$ the correlation length exponent. 
As is shown in Fig.~\ref{fig:expectations}, the theory misses most of the spectral density in the $\Delta <\omega < 4J$ range.

In this Letter, we employ quantum Monte Carlo simulations of the 2D model (\ref{BH}) in the
lattice path integral representation using the worm algorithm \cite{worm,worm2,wormLode} to study the
spectral density of the kinetic energy correlation function at zero momentum, in combination with
an analytic continuation method.
We unambiguously demonstrate the existence of a low-energy resonance peak associated
with the Higgs boson in close vicinity of the QCP by discriminating it from the second broad peak
at the typical lattice-model energies. The Higgs boson energy,
$\omega_{\rm H}$, obtained from the peak maximum increases with detuning nearly
identically as that of the particle-hole gap $\Delta_{MI}$ in the MI phase.
The spectral density associated with the Higgs boson
broadens with detuning and
quickly overlaps with other higher energy modes: It is no longer seen as a resonance
peak for a detuning as small as 20 \%, in line with the parameter regime where particle and
hole masses were found to be equal on the MI side~\cite{soyler}. On the other hand, 
in close vicinity of the QCP the Higgs boson remains visible in the spectral density at  
temperatures as high as $T_c$.  A peak is even seen in the normal phase; only at a temperature $T>2T_c$
the Higgs resonance is no longer visible. However, the onset of strong response
at low-frequency is barely modified. These results, further supported by simulations of realistic
trapped systems, explain why the experimental protocol of extracting $\omega_{\rm H}$ from the
onset of strong response \cite{bloch} works even in the absence of low-frequency resonance.

\vspace*{-2.0cm}
\begin{figure}[htbp]
\includegraphics[scale=0.4,angle=0,width=0.9\columnwidth]{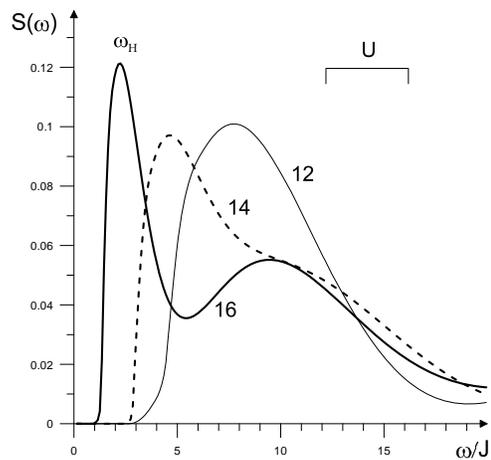}
\vspace*{-1.7cm}
\caption{\label{fig:u16to12} Spectral density $ S( \omega )$ of the kinetic energy correlation
function for $U/J=16$ (thick solid line), $14$ (dashed line), and $12$ (thin solid line)
at low temperature $T/J=0.1$. The Higgs amplitude mode ($\omega_{\rm{H}}$) emerges as a well-defined peak on approach to the quantum critical point at $U_c=16.7424$.
 }
\end{figure}

A small uniform modulation of the optical lattice depth~\cite{stoeferle} leads, under mapping to the Bose-Hubbard model (\ref{BH}),
to a perturbation proportional to the total kinetic energy of the system~\cite{kollath},
$K = - J \sum_{\langle ij \rangle}  b_i^\dag b_j^{\,} $,
\begin{equation}\label{V}
V = \delta (t) K \, , \;\;\;  \delta (t) = \frac{\delta J(t)}{J} \, ,
\end{equation}
where the small $\delta J(t)$ is proportional to the lattice modulation amplitude.
Within standard linear response theory one computes the corresponding correlation function
$\chi (i\omega_n ) = \langle K( \tau ) K(0) \rangle_{i\omega_n} + \langle K \rangle$  at
Matsubara frequencies, $\omega_n = 2\pi T n$, and performs an analytic continuation 
to obtain its spectral density $S(\omega )$. This quantity is
directly proportional to the energy absorbed by the system in the experiment~\cite{bloch}.
In the path integral representation $K(i\omega_n )$ has a straightforward Monte
Carlo estimator, $\sum_{k} e^{i\omega_n \tau_k }$, where the sum goes over all hopping
transitions in the imaginary time evolution of the system.
Thus $\chi (i\omega_n )$ is computed exactly, {\it i.e.,} the error bars
are statistical and they can be reduced arbitrarily by increasing the simulation time.
Our relative error bars for the lowest frequencies are of the order of $10^{-5}$. Nevertheless,
long simulations are required 
 because finite $\chi$ is found only after the cancelation of macroscopic factors.
The combination of the linear system size ($L=20$ in practice) of our square lattice, temperature, and $U/T$ has to be chosen such that $L$ is always significantly larger than the correlation length by a factor of at least four in order for finite size effects to be negligible. In the supplementary material we provide
details about the analytic continuation procedure which extracts the spectral density
$S(\omega) = \textrm{Im} \chi (\omega)$ from $\chi (i\omega_n )$ and show tests confirming the reliability
of the results reported here~\cite{epaps}.

\begin{figure}[htbp]
\vspace*{-0.4cm}
\hspace*{-1.5cm} \includegraphics[angle=0,width=\columnwidth]{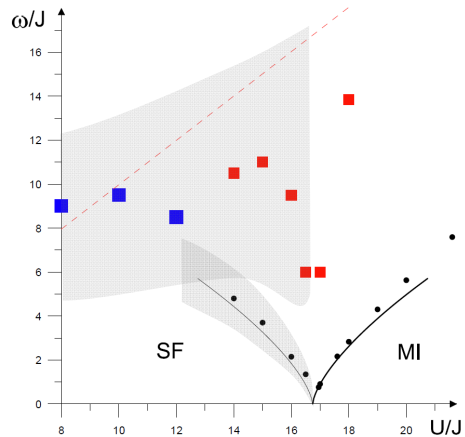}
\vspace*{-0.5cm}
\caption{\label{fig:higgs}  (Color online).
Characteristic energies in the vicinity of the quantum critical point at $(U/J)_c=16.7424$.
Black circles for $U>U_c$ and $U<U_c$ stand for particle-hole gaps
in the MI phase \cite{soyler} and energies of the Higgs bosons, respectively. Red squares
denote the location of the broad secondary peak in $S(\omega )$ until
it merges with the amplitude mode at interaction strength $U \le 12J$ to form a single peak
(blue squares). Shaded regions indicate the characteristic broadening of peaks. The thick black line
is the critical law $2.25J |(U-U_c)/J|^{\nu}$ obtained by fitting the smallest MI gaps;
its mirror reflection is shown as thin black line.
The dashed red line indicates the typical interaction scale $U/J$.
 }
\end{figure}

Unambiguous evidence for the existence of the sharp amplitude mode in the vicinity of the quantum critical
point located at $(U/J)_c=16.7424$ \cite{soyler2} is provided in Fig. ~\ref{fig:u16to12}.
We observe a well-defined resonance in the normalized spectral density at a scale much smaller
than $U$, which softens on approach to the QCP in a way that is compatible
with the $3D$ XY universality class, as is shown in Fig.~\ref{fig:higgs}.
We identify the energy of the Higgs boson
with the peak maximum. The width of the resonance peak narrows when $U \to U_c$,
suggesting that this peak is part of the universal scaling scaling function \cite{book} which can be
rewritten  as $S(\omega )\propto \omega_H^{3-2/\nu} F(\omega /\omega_H )$.
Since at frequencies $\omega_H \ll \omega $ the response must be independent of $\omega_H$
we have $S(\omega \gg \omega_H) \propto \omega^{3-2/\nu}=\omega^{0.0225}$, {\it i.e.} it is increasing
extremely slowly. The overall picture in the asymptotic limit
is that of a Higgs peak superimposed on a smeared step function, see Fig.~\ref{fig:expectations}.
When tuning away from the critical point, the Higgs mode broadens and overlaps with the second peak
around the crossover scale $U/J \approx 12$. Beyond this point the Higgs boson can no longer be discerned
as a separate mode, as is shown in Fig.~\ref{fig:u16to12}. This limits the observation of the amplitude mode to the region in close vicinity of the QCP.

\begin{figure}[htbp]
\vspace*{-1.7cm}
\includegraphics[scale=0.5,angle=0,width=0.9\columnwidth]{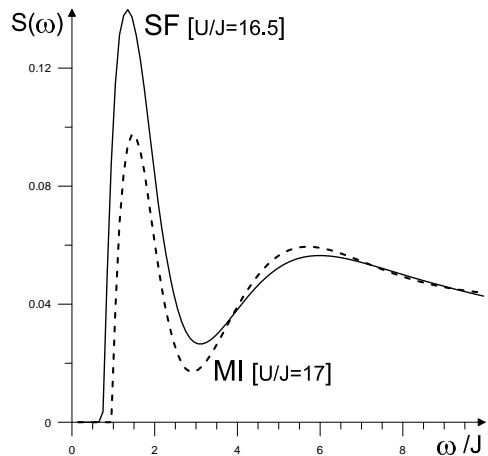}
\vspace*{-1.3cm}
\caption{\label{fig:u165u17} Spectral densities in the SF and MI phases
approximately with the same amount of detuning from quantum criticality
$|U_c-U|/U\approx 0.015$.
 }
\end{figure}

%
We also would like stress that the resonance at $\omega_H$ is seen on both sides of the
QCP, {\it i.e.} it is also present in the MI phase close to the QCP.
The similarity between the two responses for the same amount of detuning from the critical point, as is evident from Fig.~\ref{fig:u165u17}, is expected inside the correlation volume despite obvious differences in the low-frequency part. This result provides further evidence that the analytic continuation procedure is stable.

%
Though our imaginary time data decaying as $\sim 1/\tau^4$ for large $\tau$ is compatible with the scaling prediction  \cite{chubukov,sachdev} $S(\omega) \sim \omega^3$ for $\omega \to 0$, our errors bars at large imaginary times are too large and our system sizes too small to resolve this law unambiguously in analytic continuation.

There is substantial room for incoherent spectral weight between $\omega_{\rm H}$ and $\omega\approx U$ which can be filled
by other modes such as 'doublon' (double occupancy) excitations, pairs of phase modes with zero total momentum,
as well as multi-Higgs modes.
Our interpretation of the data is that higher frequency doublons ('screened' by interaction effects)
overlap with lower frequency critical phase modes creating an intermediate broad peak at frequencies between
$U$ and $zJ$, except extremely close to QCP where the second peak saturates at about $6J$ when tuning
$U \to U_c$. We associate it with pairs of phase modes with opposite momenta near the
Brillouin zone boundary (which dominate in the integral over momentum space).
This classification, however, is not rigorous in the quantum critical region
\cite{book}, as is evidenced by the similarity between the SF and MI responses.

\begin{figure}[htbp]
\vspace*{-1.0cm}
\includegraphics[scale=0.4,angle=0,width=0.8\columnwidth]{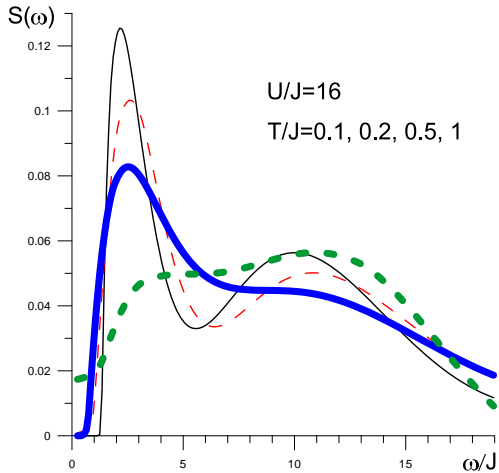}
\vspace*{-1.3cm}
\caption{\label{fig:u16T} (Color online). Evolution of spectral density with temperature at $U/J=16$.
As temperature is increased from $T/J=0.1$ (thin black line), to $T/J=0.2$ (thin dashed line), and
$T/J=0.5$ (thick blue line), the peaks get broader but remain clearly identifiable.
At  $T/J=1$ (thick dashed green line) the two peaks merge.
}
\end{figure}

Current experiments with ultra-cold atoms are typically performed at finite
temperature $T/U \ge 0.05$ (such that $T/J \sim \mathcal{O}(1)$ at $U=U_c)$ and in the presence of parabolic confinement. In Fig.~\ref{fig:u16T}
we demonstrate that for the representative case $U/J=16$ with  $T_c/J \approx 0.45$
the Higgs mode remains clearly visible at all temperatures
below the superfluid transition temperature and even slightly above it!
At a temperature $T>2T_c$ the two peaks finally merge .
Nevertheless, $S(\omega )$, still levels off at the amplitude mode frequency
$\omega \approx \omega_{\rm H}$ and has the same frequency for the onset of strong
response, meaning that these features can be used to extract $\omega_{H}$ experimentally.
Note that phase coherence can extend across finite systems
at temperatures well above the thermodynamic $T_c$ for the
Kosterlitz-Thouless transition  characterized by an exponentially divergent correlation length.

\begin{figure}[htbp]
\vspace*{-1.0cm}
\includegraphics[scale=0.4,angle=0,width=0.8\columnwidth]{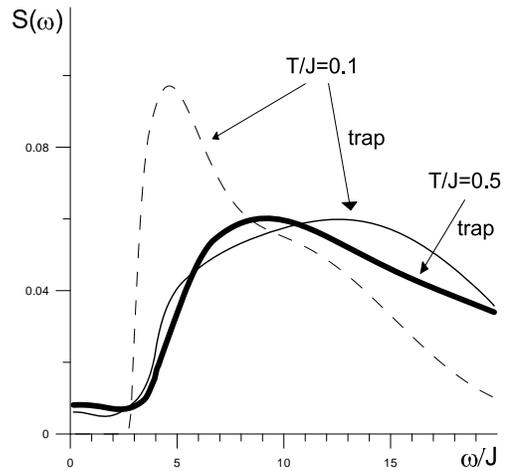}
\vspace*{-1.3cm}
\caption{\label{fig:trap} Effects of trapping potential and finite temperature
for $U/J=14$ ($T_c/J \approx 1.04$). The spectral density of a homogeneous system at low temperature with Higgs resonance (dashed line) transforms into a broad (irregular) peak due to inhomogeneous broadening in a trapped system
with $N=190$ particles at unit filling factor in the middle (thin solid line). At a temperature
$T/J=0.5$ (thick solid line) we observe a smooth single peak.
}
\end{figure}

Inhomogeneous broadening of spectral density caused by the trap has a dramatic effect
on the structure of $S(\omega )$ as signals from different parts of the system are superimposed on each other. Moreover,
in the presence of external potential gradients the spectral density is no longer vanishing at
 $\omega \to 0$ because of low-frequency sound modes (predominantly in the trap edges),  
 in line with experimental observations~\cite{bloch}.
Under these conditions, the Higgs mode can no longer be seen as a sharp resonance in $S(\omega )$. There is instead a broad maximum with irregular shape. Finite temperature effects further
transform it into a smooth single peak. In Fig.~\ref{fig:trap} we show the comparison between
the homogeneous and trapped cases. 
The simulation was performed for realistic experimental parameters \cite{bloch} 
but at a variety of 
temperatures in order to discriminate between trap and temperature effects. Even though the resonance is
lost in the total signal, the steep onset of spectral response still correlates remarkably well
with the energy of the Higgs boson.

In conclusion, we find that the Higgs boson is  a well-defined though significantly damped excitation
 in close vicinity ($\sim 20\%$) of the particle-hole
symmetric and Lorentz-invariant SF-MI quantum critical point in two dimensions.
It is seen as a resonance in the kinetic energy correlation function which is directly probed through
the modulation of the optical lattice depth in experiments with ultra-cold atoms.
The energies of the amplitude mode match particle-hole gaps in the Mott insulator phase
for the same amount of detuning away from quantum criticality.
While temperatures at least as high as the critical temperature for superfluidity
preserve the Higgs resonance, inhomogeneous broadening in small trapped systems
erases resonance-type features in the spectral function. Nevertheless, it is possible
to determine the energy of the amplitude mode from the onset of strong response, as is done
in a recent experiment \cite{bloch}.

This work was supported by the National Science Foundation under grant PHY-1005543.
Use was made of the maxent application~\cite{maxent} in the ALPS libraries~\cite{alps2}.
We thank S. Sachdev, B. Svistunov, W. Zwerger, and the authors of Ref.~\cite{bloch} for insightful discussions, as well as for sharing their results with us prior to publication.


\vspace*{2cm}

{\huge{Supplementary material for ``The Higgs mode in a two-dimensional superfluid"}}






\section{Quantum Monte Carlo simulations}


\begin{figure}[t]
\vspace*{-2.cm}
\includegraphics[angle=0,width=1.0\columnwidth]{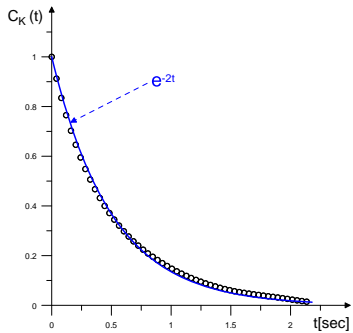}
\vspace*{-2.5cm}
\caption{\label{fig_S0}  (Color online).
Kinetic energy autocorrelation function for $L=20$, $T/J=0.1$, and $U/J=16.5$ (black circles)
described by the near exponential decay with integrated autocorrelation time $\tau_A \approx 0.5 \,sec$.
}
\end{figure}
We simulate the two-dimensional Bose-Hubbard model
\begin{equation} 
H = - J \sum_{<ij>}  b_i^\dag b_j^{\,}
+ \frac{U}{2} \sum_i n_i(n_i-1) - \sum_i (\mu -v_i)n_i  \, ,
\end{equation}
with  hopping amplitude $J$, on-site interaction $U$, chemical potential $\mu$, and trapping potential $v_i$
using quantum Monte Carlo simulations in the path-integral representation with worm-type updates~\cite{worm,worm2,wormLode}. Unless otherwise stated, our unit is the hopping $J=1$.
The worm algorithm is a well established method in the study of bosonic systems such as cold gases and Helium-4. It can not only simulate bosons loaded in an optical lattice realistically, but even go beyond that and study a million particles for temperatures of the order of the hopping on a single CPU.
It is well established that Worm Algorithm does not suffer from the critical slowing down problem and remains efficient on approach to the critical point. For our system size at the lowest simulated temperature and at U=16.5 (closest to the quantum critical point) the measured autocorrelation time for kinetic energy at the lowest frequency (the slowest mode) was about 0.5 sec (see Figure \ref{fig_S0}). With the typical simulation time exceeding three weeks on a single CPU this translates to $>$ 3 million uncorrelated samples. Note also, that our space-time volume in this case
contains about 200 correlation length volumes.

We consider the response of the system to a small uniform modulation of the optical lattice depth, which can be described by a perturbation $V$ which is proportional to the total kinetic energy of the system, $K = - J \sum_{<ij>}  b_i^\dag b_j^{\,} $, namely
\begin{equation}
V = \delta (t) K \, , \;\;\;  \delta (t) = \frac{\delta J(t)}{J} \, ,
\end{equation}
where $\delta J(t)$ is proportional to the lattice modulation amplitude.
We consider the response to lowest non-trivial order when the response is linear in time and quadratic in the perturbation amplitude, which is valid for weak perturbations under otherwise adiabatic conditions of the system.
Experimentally, this quantity is directly proportional to the energy absorbed by the system when modulation at different frequencies is performed for a fixed number of cycles~\cite{bloch}.

In this regime, we are allowed to consider the correlation function
\begin{equation}
\chi (i\omega_n ) = \langle K( \tau ) K(0) \rangle_{i\omega_n} + \langle K \rangle,
\end{equation}
  at Matsubara frequencies $\omega_n = 2\pi T n$ ($n>0$), compute it numerically and perform an analytic continuation procedure to obtain its spectral density $S(\omega )$. The subscript ${i\omega_n} $ denotes that the Fourier transform is taken of the corresponding correlation function $\langle K( \tau ) K(0) \rangle$ in imaginary time.
In our simulations, the Fourier transform is taken on the flight, {\it i.e.,} we immediately collect statistics for different Matsubara frequencies, which can be done much faster than for the correlation function in imaginary time.
The quantity $\chi(i\omega_n)$ approaches a constant for large Matsubara frequencies, namely minus the kinetic energy $-\langle K \rangle$. Statistics for the kinetic energy are collected separately, and the quantity is afterwards subtracted from the measurements for the correlation function. The correlation function at zero frequency has to be discarded because it contains no information on the Higgs mode,  and is orders of magnitude larger than the signal at non-zero Matsubara frequencies.
We considered system sizes $L\times L = 20 \times 20$ and inverse temperatures $\beta \le 10$. The correlation length, determined from the one-body density matrix,  scales as $\xi_c/a \approx 5 J/ E$ where $E$ is the characteristic energy scale of the problem
($T$ at the critical point~\cite{soyler} $U_c = 16.7424$, particle-hole gap $\Delta_{MI}$ in the MI phase, or the Higgs mode frequency
in SF) with $a$ the lattice spacing.
 Except in very close vicinity of the critical point, $\vert U - U_c \vert < 0.2$, our system size is thus much larger than the correlation length.
 Since measuring correlation functions is one to two orders of magnitude more difficult than single particle properties, the increase in CPU would be too expensive to study larger system sizes, while not altering our conclusions.

\section{The data at large imaginary frequencies}

\begin{figure}[htbp]
\includegraphics[angle=0,width=0.8\columnwidth]{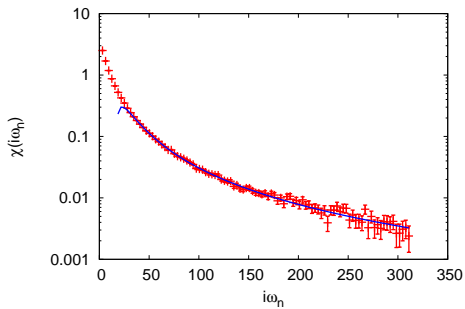}
\caption{\label{fig_S1}  (Color online).
Monte Carlo data for the correlation function $\chi(i\omega_n)$ as a function of the Matsubara frequencies $i\omega_n$. The solid line is a fit of the form $f(x) = \frac{a}{x^2} + \frac{b}{x^4}$, which is used for $i\omega_n > 50$. Parameters are $J=1$, $\beta = 2$, $U=14$ and $\langle n \rangle = 1$.
 }
\end{figure}

\begin{figure}[htbp]
\includegraphics[angle=0,width=0.8\columnwidth]{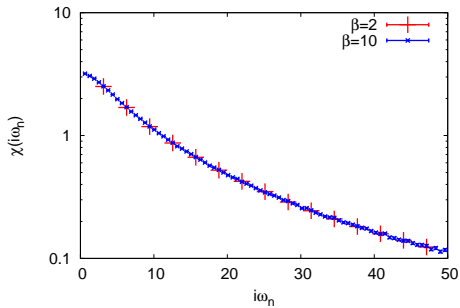}
\caption{\label{fig_S2}  (Color online).
Monte Carlo data for the correlation function $\chi(i\omega_n)$ as a function of the Matsubara frequencies $i\omega_n$ for two different inverse temperatures, $\beta = 2$ and $\beta = 10$. Error bars are shown, but are very small. Same parameters as in Fig.~\ref{fig_S1}.
 }
\end{figure}

A typical set of Monte Carlo data is shown in Fig.~\ref{fig_S1}. The data for large Matsubara frequencies are fitted using the form $f(x) = \frac{a}{x^2} + \frac{b}{x^4}$, with $a,b$ fitting parameters. Only even powers of $\omega_n$ are allowed by causality and symmetry. The fit is used for frequencies $i\omega_n > 50$. At low temperatures, say $\beta = 10$, the high frequency tail of the correlation function is indistinguishable within the error bars from the one at a lower temperature $\beta = 2$ (see Fig.~\ref{fig_S2}). We use therefore the tail determined at $\beta=2$ for the $\beta=10$ case as well, hereby substantially reducing the required CPU at low temperatures (we measure at most the correlation function for a 100 Matsubara frequencies at every temperature).

\begin{figure}[htbp]
\includegraphics[angle=0,width=0.8\columnwidth]{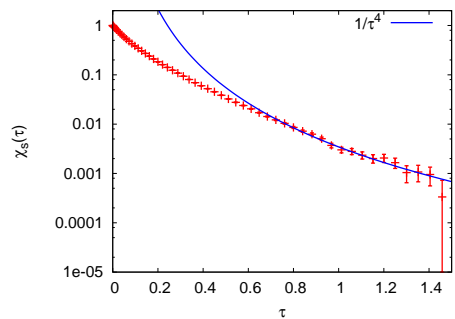}
\caption{\label{fig_S3}  (Color online).
The correlation function $\chi_s(\tau)$ as a function of imaginary time $\tau$ over the range $\tau \in [0, 1.5]$. For large $\tau$, the data are compatible with the asymptotic behaviour $1/\tau^4$, corresponding to $S(\omega)\sim\omega^3$ for small $\omega$, predicted by scaling \cite{chubukov,sachdev}.
Parameters are $J=1$, $\beta = 10$, $U=14$ and $\langle n \rangle = 1$.
 }
\end{figure}

The next step is applying a Fourier transform to the correlation function in Matsubara domain to the imaginary time domain. Because of the slow convergence of the functions $1/(i\omega_n)^2$, we sum up to 500,000 frequencies to ensure that frequency truncation has no detectable effect.
This correlation function in imaginary time approaches, in general, a non-zero constant for $\tau \to \beta/2$. This non-zero constant will be seen as a delta-peak at $\omega=0$ for $S(\omega)$. We subtract this delta-functional contribution from $\chi(\tau)$ by shifting the data
resulting in the correlation function $\chi_s(\tau)$, which is shown in Fig.~\ref{fig_S3}.
For large $\tau$, this function can be fitted to $\sim 1/\tau^4$, in line with theoretical predictions  \cite{chubukov,sachdev}
for the scaling behavior of the spectral function $S(\omega) \sim \omega^3$ at $\omega \to 0$, though our errors bars in the tail
are too large to resolve this law unambiguously.

\section{Analytic continuation}

\begin{figure}[htbp]
\includegraphics[angle=0,width=0.8\columnwidth]{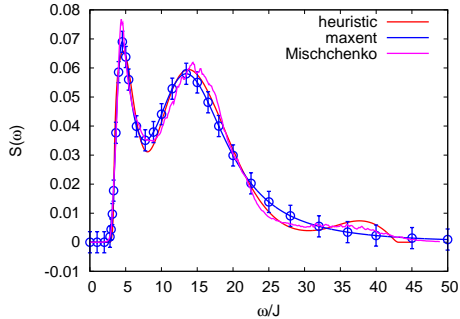}
\caption{\label{fig_S4}  (Color online).
Spectral density $S(\omega)$ obtained by different analytic continuation methods. Parameters are $J=1$, $U=18$, $\beta=1$, and $\langle n \rangle = 1$. The gap is $\Delta/J = 2.8$, see Ref.~\onlinecite{soyler}, in agreement with the onset of spectral weight in this figure. The maxent parameters~\cite{maxent,alps2} are $\chi^2=0.50$ and $\alpha = 0.346$ with a flat default model. Error bars determined by ALPS~\cite{alps2} on the maxent curve for a few data points are shown.
 }
\end{figure}

We are now in a position to perform analytic continuation. The spectral density $S(\omega) \ge 0 (\omega \ge 0)$ is related to the correlation function in imaginary time $\chi_s(\tau)$ by
\begin{equation}
\chi_s(\tau) = \int d\omega K(\omega, \tau) S(\omega),
\end{equation}
with the kernel
\begin{equation}
K_0(\omega, \tau) = \exp(-\omega\tau),
\end{equation}
at zero temperature, and
\begin{equation}
K(\omega, \tau) = \exp(-\omega\tau) + \exp(-\omega(\beta - \tau)),
\end{equation}
at finite temperature $T=1/\beta$. Given $S(\omega)$, the computation of $\chi_s(\tau)$ is easy. The inverse problem is equivalent to an inverse Laplace transform, which is exponentially sensitive to the data. Only when the function $\chi_s(\tau)$ is known analytically over a finite interval, analytic continuation can be done reliably (analytically), but in the presence of statistical noise and grid binning errors, the problem is ill-defined.

Practical schemes seek a compromise between smoothening the target function and reproducing $\chi_s(\tau)$ within error bars. Smoothening is necessary to acquire a physically acceptable solution.  The degree and type of smoothening is never independent of the choices made by the practitioner or the used software. However, the better the quality of the data in imaginary time, the more acceptable solutions are constrained, and hence the better the analytic continuation can be performed. This renders a correct error analysis indispensable, which was done here using the alea library of the ALPS project~\cite{alps2}. On general grounds, one expects that gaps, the position of the first peak and the integrated weight are reliable, but that high frequency features are beyond control, and that those may have a back-action on the width of the peaks.

We applied three analytical continuation methods to our data in order to extract what features are seen by different methods, and are hence likely to be physical. We used maxent~\cite{maxent} in the ALPS implementation~\cite{alps2}
, stochastic inference~\cite{beach,fuchs} in the implementation of Ref.~\cite{mishchenko},
and a novel heurisitic procedure which seeks the best solution reproducing $\chi_s(\tau)$ with additional
self-consistently adjusting quadratic measures favoring smooth non-negative functions.
The comparison is shown in Fig.~\ref{fig_S4}. We see that the position of the first and the second peak are reliable,
just as the integrated weight of the spectral density.
We do not report any data at frequencies higher than $20J$ in this article.
Also till what $\tau$ the function $\chi_s(\tau)$ can be computed matters enormously.


\section{Comparison with Ref.~\cite{pod11}}

\begin{figure}[htbp]
\includegraphics[angle=0,width=0.8\columnwidth]{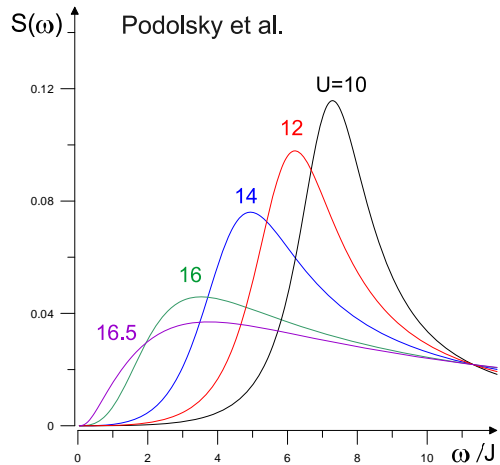}
\caption{\label{fig_S5}  (Color online).
Spectral density of the scalar susceptibility according to Ref.~\cite{pod11} for different values of $U/J$ based on a large
$N=2$ expansion in $d=2$ dimensions. At low $\omega$, the behavior of the scalar susceptibility is $~\omega^3$.
The amplitude was chosen arbitrarily because of the high frequency tail $\sim 1/\omega$.
The theory predicts a sharpening of the peak when $U$ is decreased, and saturation of the response in the maximum region
when $U \to U_c$, i.e. no softening is observed at criticality.
}
\end{figure}

\begin{figure}[htbp]
\includegraphics[angle=0,width=0.8\columnwidth]{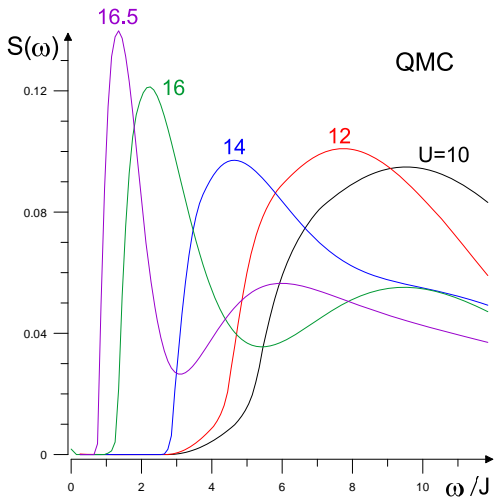}
\caption{\label{fig_S6}  (Color online).
Spectral density $S(\omega)$ for low frequencies obtained by Monte Carlo and analytic continuation for different values of $U/J$.
This should be compared with Fig.~\ref{fig_S5} (see also text).
 }
\end{figure}

\begin{figure}[htbp]
\includegraphics[angle=0,width=0.8\columnwidth]{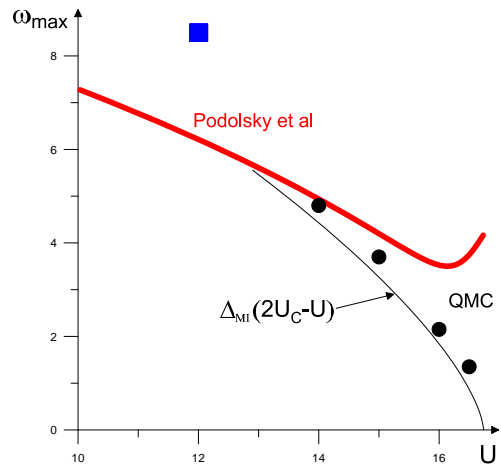}
\caption{\label{fig_S7}
(Color online). Comparison of the peak maximum between QMC (black circles for Higgs mode and blue square for the broad
peak at $U/J=12$) and Eq.~\ref{eq:pod} (red line). Also shown is the gap determined in the Mott Insulator from the single particle Green function~\cite{soyler} by mirror reflection around the critical point (thin black line). The maximum agrees for $U/J=14$, which is where QMC predicts the Higgs mode to be only barely discernible. For larger values of $U$ the peak maximum
in Eq.~\ref{eq:pod} saturates to the value predicted by $\gamma$ alone, i.e. in the critical region the peak maximum in Eq.~\ref{eq:pod}
is not linked to the Higgs mass.
 }
\end{figure}

The theory of Ref.~\cite{pod11}
predicts for the imaginary part of the scalar susceptibility $\chi_{\rho \rho}''(\omega)$ (to be identified with our $S(\omega)$),
\begin{eqnarray}
\chi_{\rho \rho}''(\omega) & \propto & \frac{U^2\omega^3}{(\omega^2 - m^2)^2 + 4 \gamma^2 \omega^2}, \label{eq:pod} \\
\gamma & = & U/8J, \\
m & = & 8\sqrt{2} J \, (1-U/U_c)^{1/2}
\end{eqnarray}
In this large $N$ expansion, $N=2$ for $<n>=1$ were set. This parametrization follows from App. A in Ref.~\onlinecite{polkovnikov} (involving a large occupation number Ansatz. Also a constant density of states for the phonons is assumed.) and was also used in the analysis of the experiment~\cite{bloch}. At low frequencies $\chi_{\rho \rho}''(\omega) \sim \omega^3$, in line with Ref.~\cite{chubukov}.
At high frequencies, $\chi_{\rho \rho}''(\omega) \sim \omega^{-1}$, a consequence of the truncation scheme in the theory.
This means, in particular, that the normalization integral diverges and thus can be chosen arbitrarily for comparison with
simulation.

The crossover from $\omega^3$ to $\omega$  behavior at low frequencies in Eq.~(\ref{eq:pod}) occurs at
frequencies $\sim m^2/2\gamma \sim (1-U/Uc)$ and constitutes strong violation of the universal low-frequency
scaling law, $S(\omega )\propto \Delta^{3-2/\nu} F(\omega /\Delta )$ (with $\Delta \sim (1-U/U_c)^{\nu}$ and
the correlation length exponent $\nu=0.6717$), obtained by dimensional analysis \cite{book}.
It predicts that after crossover at $\omega \sim \Delta$ the response function increases as
$S(\omega ) \sim \omega^{3-2/\nu}=\omega^{0.0225}$.

The (unnormalized) response is shown in Fig.~\ref{fig_S5}, and should be compared with the (normalized) response computed by quantum Monte Carlo simulations shown in Fig.~\ref{fig_S6}.  Equation~\ref{eq:pod} has the following behavior,
\begin{itemize}
\item For $2\gamma < m$  (this corresponds to $U/J > 15$), the response is peaked at the Higgs mass.
      The position of the maximum is within the peak width the same as the one found in Fig.~\ref{fig_S6}.
      However, the response becomes narrower for smaller values of $U/J$, unlike the Monte Carlo prediction in Fig.~\ref{fig_S6}.
\item For $2\gamma > m$ (this corresponds to $15 < U/J < U_c/J$), the response saturates when tuning closer to the critical point. No further softening can be seen, only a broad response, where the maximum can no longer be identified with the Higgs mass. Note that the QMC behavior in this regime is different (see Fig.~\ref{fig_S6}):
    further softening and narrowing of the low-frequency Higgs resonance is observed suggesting that this peak is part of the universal scaling
    scaling function, and a second peak emerges at $\omega/J \approx 6$ (the latter is not part of the theory of Ref.~\onlinecite{pod11}).
\end{itemize}
A comparison for the peak maximum is shown in Fig.~\ref{fig_S7}.

\section{Disappearance of the Higgs resonance in the Mott phase}

\begin{figure}[htbp]
\vspace*{-2.cm}
\includegraphics[angle=0,width=1.0\columnwidth]{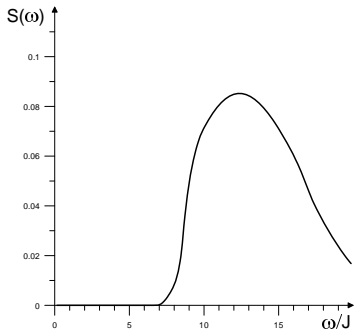}
\vspace*{-2.cm}
\caption{\label{fig_S8}
Spectral density in the Mott phase at $U/J=20.5$.
 }
\end{figure}
On the superfluid side of the transition, our data indicate that the Higgs resonance is seen only 
in close vicinity of the transition point (within 20\%). The control simulation of the MI phase 
at $U/J=20.5$ confirms that this 20\% figure of merit can be used on both sides of the transition, 
see Fig.~\ref{fig_S8}.

\section{Simulations in a trap}

In the experiment~\cite{bloch}, a particle number of $\langle N \rangle = 190(36)$ was reported. Together with the requirement that the density is one in the trap center, this corresponds to a trap parameter $v_c(x,y) = 0.0915 ((x-x_c)^2 + (y-y_c)^2)$ in units of the hopping for $U=14$. Here, $(x_c, y_c)$ are the lattice coordinates of the trap center. The reported temperatures $T/U \sim 0.1$ correspond to an entropy $S/N = 0.8$ in the atomic limit. Assuming adiabaticity at all stages in experiment, this would mean that $\beta J= 0.6$ in the trap for $U=14$, which is a temperature 50\% above $T_c$. Since the passage to the atomic limit is not adiabatic, and that substantial energy is added during the 20 cycles of the lattice modulation (also due to quantum optics effects, as was seen in Ref.~\cite{trotzky}, the true temperature before the lattice modulation is lower, perhaps by  50\%. The heating has not been addressed experimentally, and therefore a more quantitative estimate of temperature is not possible. The rule of thumb $T/U \sim {\rm const.}$ for presently realistic temperatures of the scale of the bandwidth and for values of $U$ outside the Bogoliubov regime, nevertheless holds, in agreement with previous experiments~\cite{greiner, kuhr}. \\

\end{document}